\documentclass[final,5p,times,twocolumn]{elsarticle}
\usepackage{graphicx} 
\usepackage{orcidlink}
\usepackage{siunitx}
\usepackage[version=4]{mhchem}
\usepackage[utf8]{inputenc}
\usepackage{cleveref}
\title{From Single Particles to Clinical Beam Rates: A Wide Dynamic Range Beam Monitor}

\author{Simon Waid}
\author{Philipp Gaggl}
\author{Andreas Gsponer}
\author{Richard Thalmeier}
\author{Jürgen Burin}
\author{Matthias Knopf}
\author{Thomas Bergauer}

\address{Institute of High Energy Physics, Austrian Academy of Sciences, Nikolsdorfergasse 18, 1050 Wien, Austria}

\journal{NIM A}

\begin{document}

\begin{abstract}
Access to high-energy particle beams is key for testing high-energy physics (HEP) instruments. Accelerators for cancer treatment can serve as such a testing ground. However, HEP instrument tests typically require particle fluxes significantly lower than for cancer treatment. Thus, facilities need adaptations to fulfill both the requirements for cancer treatment and the requirements for HEP instrument testing.  We report on the progress made in developing a beam monitor with a sufficient dynamic range to allow for the detection of single particles, while still being able to act as a monitor at the clinical particle rates of the MedAustron treatment facility. The beam monitor is designed for integration into existing accelerators. 

\end{abstract}
\maketitle

\section{Introduction}
The MedAustron accelerator in Wiener Neustadt was designed as a cancer treatment and research facility. However, the beam monitors installed initially in the facility are only suitable for particle rates used in cancer treatment, which for proton beams are typically between $10^9$~\unit{\per\second} and $10^{11}$~\unit{\per\second}. The beam monitors cannot resolve the beam properties at the particle rates commonly used for HEP instrument testing (kHz to MHz particle rates). Nevertheless, beams with particle rates down to 3 kHz were commissioned for research purposes \cite{ulrich-purCommissioningLowParticle2021} and are available at the facility. This commissioning, however, was carried out purely using optics simulations and an iso-center monitor. 

Beam monitors covering the full range of available particle rates would further benefit quality assurance and future commissioning tasks. We previously investigated different options for implementing such a monitor \cite{waidSiCBasedBeam2024, waidDetectorDevelopmentParticle2024}. The most promising option was to measure and integrate charge from a DC-coupled strip sensor for fixed time intervals. The charge in a given time interval is proportional to the number of particles. Single particles can be detected with sufficiently high charge resolution and low noise. The detector can operate at clinical particle rates, provided the readout system has sufficient dynamic range. 

At clinical rates, the beam is quasi-continuous. Thus, DC coupling is required.  The natural material choice for the sensor would be silicon(Si). However, Si sensors exhibit substantial dark current when radiation-damaged by ion beams. Given the requirement of DC-coupling, Si can not be used as a sensor material. Instead, we use a silicon carbide (SiC) strip sensor, for which we observed a dark current of less than \qty{1}{\pico\ampere\per\square\centi\meter} even if radiation damaged with doses up to $10^{15}$ $\mathrm{n_{eq} \thinspace cm^{-2}}$ \cite{waidDetectorDevelopmentParticle2024}. This enables the detection of charges generated by single ions, even with a DC-coupled sensor \cite{waidSiCBasedBeam2024, waidDetectorDevelopmentParticle2024}.  

The implemented sensor readout is based on the Analog Devices AD8488 front-end \cite{waidSiCBasedBeam2024}. We have previously shown that the noise levels of the front-end can reach \qty{600}{\elementarycharge} \cite{waidSiCBasedBeam2024, waidDetectorDevelopmentParticle2024}. When coupled with a sufficiently thick sensor, this enables single-particle detection. To attain a desired signal-to-noise ratio (SNR) of 10 when coupled with SiC sensors having a thickness of at least \qty{100}{\micro\meter}. Based on our previous findings, we present an upscaled version of the detector readout system, details on the sensor design, and results of beam tests. 

\section{System Design}

\begin{figure}
    \centering
    \includegraphics[width=1\linewidth]{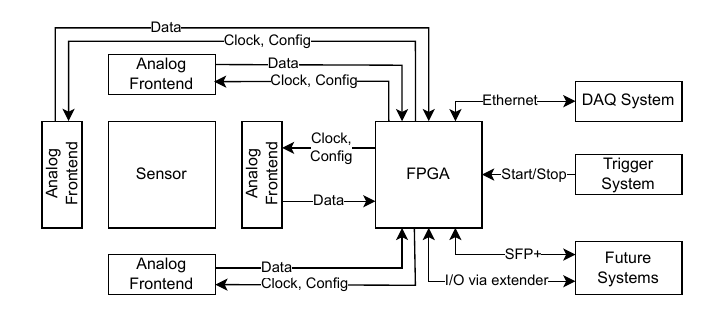}
    \caption{System architecture of the beam monitor. Four analog frontends are controlled via a Zynq Ultrascale+ MPSoC FPGA/CPU combination under the black heat sink on the right-hand side of the image. The FPGA part controls the analog front end and the trigger system, while the CPU communicates with the DAQ system. Additional interfaces are available for interfacing with other systems, including real-time data transfer via SFP+.}
    \label{fig:architecture}
\end{figure}

The overall system architecture of the upscaled beam monitor is shown in \cref{fig:architecture}. At the sensor's core is a Zynq Ultrascale+ MPSoC, which is a combination of a freely programmable gate array (FPGA) and a CPU. The FPGA part performs real-time tasks, such as generating the clocks for the analog frontends and reading back ADC values. A Linux-based operating system on the CPU provides functionality, such as a TCP/IP stack and kernel drivers for interfacing the FPGA, simplifying software development. A program written in C and running in user space communicates with the DAQ system and configures the hardware. Data exchange between the CPU and FPGA utilizes direct memory access (DMA) transfers to offload the CPU. Given that the system generates 33 MBytes per second of data when running at full speed, a compression algorithm for the data was implemented. Using delta compression, the data rate is approximately halved. Depending on the amplification and filter settings, as well as on the presence of the beam, we observed practical data rates between 15 and 20 MBytes per second. The compression is performed on the FPGA, minimizing CPU usage.

Data is transferred in variable-sized packages to a DAQ system. For data transfer, Gigabit Ethernet is employed. Each data package contains 16 analog frontend samples, time stamps, and a checksum for integrity verification. Trigger inputs are provided to allow for synchronization with the accelerator. Upon detection of a trigger event, the timestamp of the trigger event is recorded and sent to the DAQ system via a dedicated package in the data stream. Typically, the accelerator provides two trigger signals, indicating the beginning and end of a measurement. Trigger events can be used to reduce the congestion of the network and the sampling rate. This way, a lower data rate can be attained when no active measurement is running.

\begin{figure}
    \centering
    \includegraphics[width=1\linewidth]{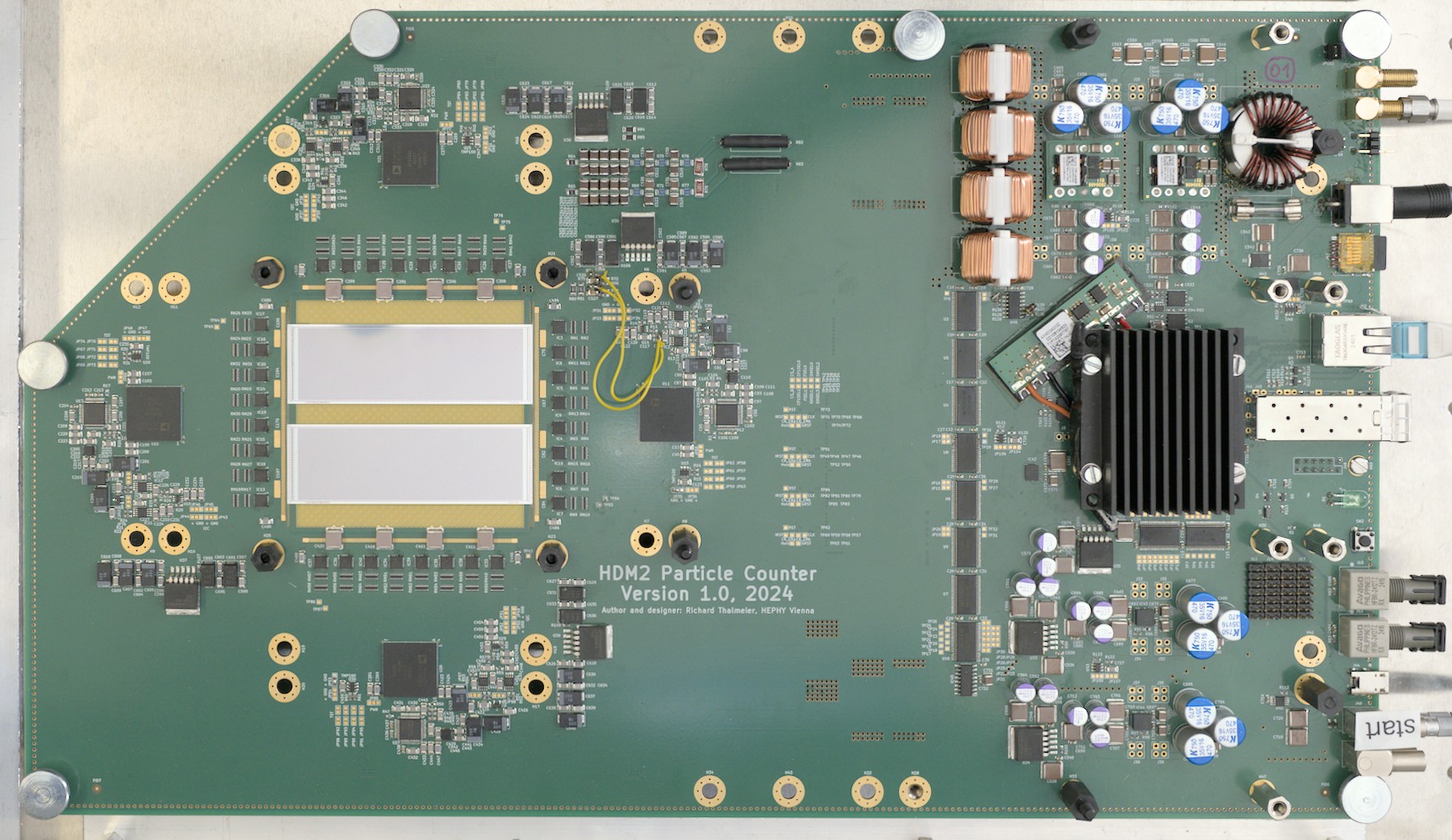}
    \caption{Photograph of the prototype system undergoing beam tests. Due to delays in the production of SiC sensors, Si strip sensors were employed for beam testing. }
    \label{fig:foto}
\end{figure}

A photograph of the sensor and readout part of the system is shown in \cref{fig:foto}. On the left side, the sensor is visible. At the moment of writing, the SiC sensors designed for the monitor were not yet available. Instead, two Si strip sensors are mounted, leaving gaps in the active area. Around the sensor are four analog frontends, each composed of an AD8488 integrator array and an AD9244 ADC. Each direction of strips is read out by analog frontends.  The FPGA controls the clock of each frontend individually and can read out all frontends in parallel. Strips of each direction are alternately connected to the front-ends, partitioning them into even and odd strips. Even strips are connected to one front-end, and odd strips to the other front-end. Given that the clocks of the analog frontends can be driven individually, the integration/readout of even and odd strips can be shifted in time, allowing for an interleaved readout. This interleaved operation allows for higher sampling rates at the cost of lateral resolution. In synchronous mode, the maximum sampling rate is \qty{37}{\kilo\hertz}, while in interleaved mode it is increased to \qty{70}{\kilo\hertz}. Besides a Gigabit Ethernet interface, a small form-factor pluggable plus (SFP+) cage for real-time data transfer is provided. An additional I/O expander can be stacked on top of the PCB (not shown). This allows for future extension of the functionality of the beam monitor.

\section{Sensor}

\begin{figure}
    \centering
    \includegraphics[width=0.9\linewidth]{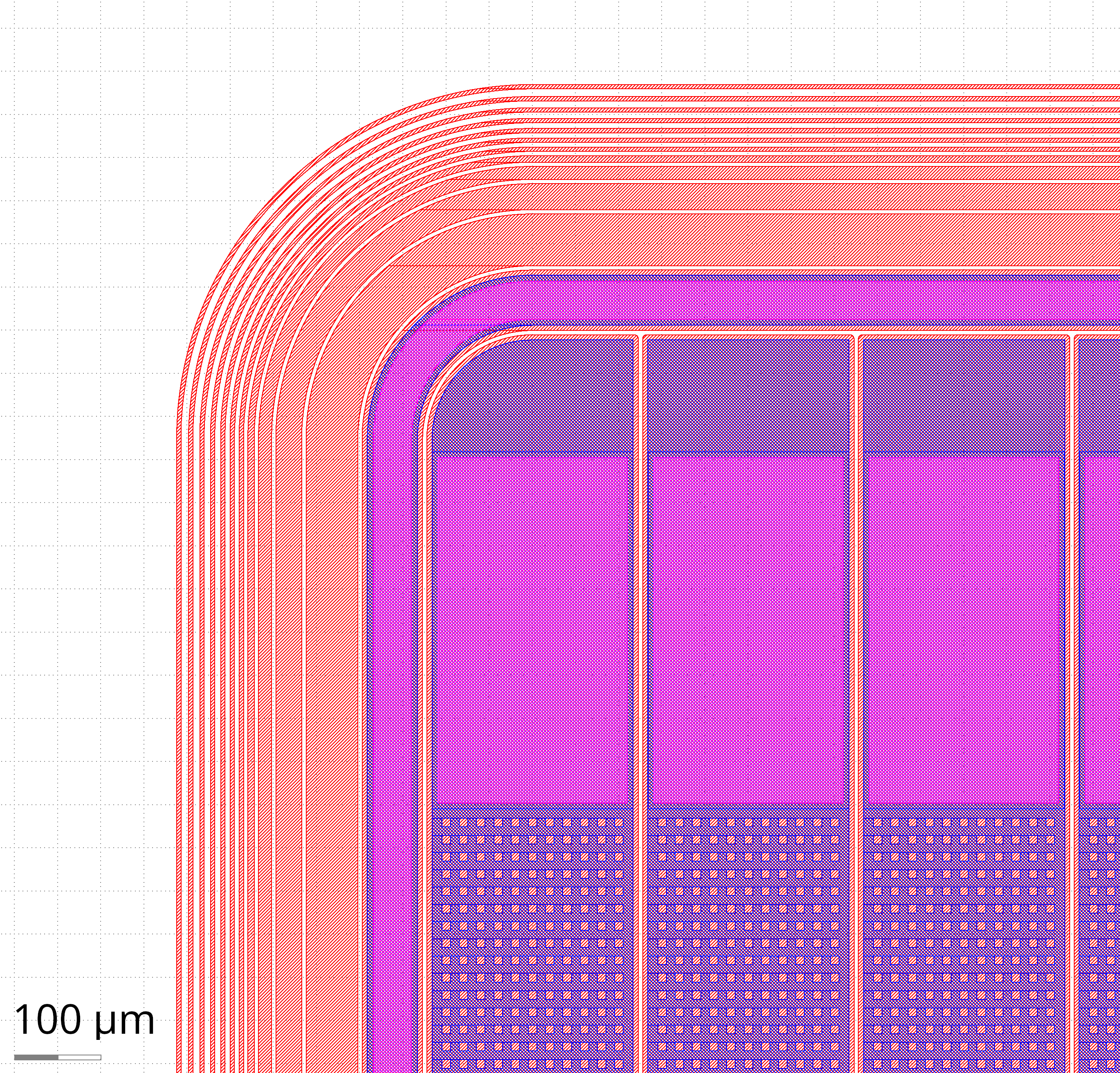}
    \caption{Closeup on the high-voltage optimized radiation SiC-sensor design. The p-implant is drawn in red; blue shows the metallization, and the passivation openings are pink.}    
    \label{fig:sensor}
\end{figure}

We previously found that the charge collection efficiency (CCE) of radiation-damaged SiC scales with the applied reverse voltage. At sufficient depletion voltages, operation of SiC sensors exposed to $5 \cdot 10^{15}$~$\mathrm{n_{eq}/cm^2}$ is feasible with a CCE above 60\% \cite{gagglChargeCollectionEfficiency2022}. Thus, we optimized the sensor design for high reverse voltages, well above the expected depletion voltage. This optimization was performed using the Synopsys Sentarus technology computer-aided design (TCAD) software. A sensor thickness of \qty{100}{\micro\meter} was chosen. The design of the sensor is shown in \cref{fig:sensor}. It uses a p-implanted guard ring structure to distribute and degrade fringe fields responsible for premature device breakdown. This guard structure was tailored for maximum operating voltage. Due to expected yield limitations during sensor production, we plan to manufacture large-area sensors out of smaller sensor tiles. To keep the blind area between sensors low, a second optimization criterion for the guard rings was low-area usage. Further, the design is tailored for manufacturing using a full-wafer shadow mask. Therefore, the minimum feature size of the design is \qty{5}{\micro\meter}. The guard ring structure consists of an initial collector ring, which can be contacted to minimize leakage currents. The other rings gradually degrade fringe fields towards the outside. The collector ring has a width of \qty{75}{\micro\meter} and could be omitted to save space at the cost of larger dark current on the sensor strips. Further, doing so would be detrimental to CV measurements commonly employed in sensor characterization. The fringe field degrading rings have a total width of \qty{250}{\micro\meter}. The sizes of the individual rings are detailed in \cref{tab:guardrings}. Simulations indicate the device may be operated at up to \qty{4.5}{\kilo\volt}.

\begin{figure*}
    \centering
    \includegraphics[width=0.9\linewidth]{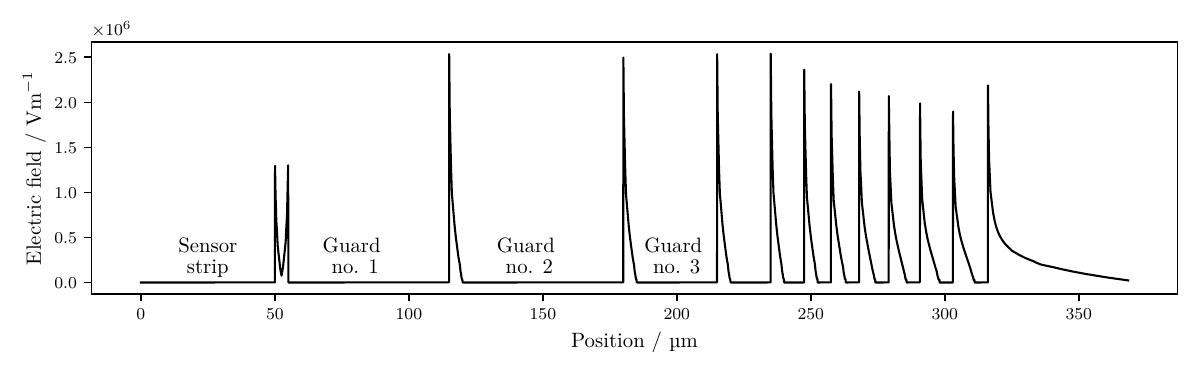}
    \caption{Simulated field intensities along the guard rings at 4553V, close to breakdown. In this case, the first guard ring is contacted at the sensor bias.}
    \label{fig:field}
\end{figure*}

The simulated field intensities close to breakdown (around 2.7 MV/cm) along the sensor are shown in \cref{fig:field}. From left to right, \cref{fig:field} shows the fields at the border of the active area, as well as at the border of each guard ring. The guard rings were optimized to distribute the potential degradation, and the resulting field peaks uniformly across all guard rings while simultaneously minimizing space usage. The simulation also considered the insulating layer stack of \qty{700}{\nano\meter} \ce{SiO2} and \qty{500}{\nano\meter} \ce{Si3N4} above the SiC for passivation and potential breakdown therein. Results suggest the possibility of substantial electric fields at the passivation surface. To prevent ionization of the surrounding air at high operating voltages, the sensor must be covered with an insulating material capable of withstanding the electric field, such as siloxane polymers. For sensor probing using optical methods, such as the transient current technique (TCT), the metal covering the implant features a grid pattern with openings. The holes in the metal grid are 10x10 \unit{\square\micro\meter} in size and cover one-quarter of the area. 

\begin{table}
    \centering
    \begin{tabular}{lrr}
        Ring No. & Ring width (µm) & Spacing (µm)\\
        1 & 75.0 & 5.0 \\
        2 & 60.0 & 5.0 \\
        3 & 30.0 & 5.0 \\
        4 & 15.0 & 5.0\\
        5 & 7.5 & 5.0\\
        6 & 5.0 & 5.5\\
        7 & 5.0 & 6.0 \\
        8 & 5.0 & 6.6 \\
        9 & 5.0 & 7.3 \\
        10 & 5.0 & 8.0 \\
        11 & 5.0 & 8.9 \\
        12 & 5.0 &  \\         
         
    \end{tabular}
    \caption{High voltage optimized guard rings design. The numbering starts with the ring closest to the sensor. The spacing refers to the distance towards the next ring, away from the sensor. Ring no. 1 can be contacted to minimize surface leakage currents and enable precise CV measurements. If this functionality is not needed, it can be omitted.}
    \label{tab:guardrings}
\end{table}

\section{Measurement Results}

\begin{figure}
    \centering
    \includegraphics[width=1\linewidth]{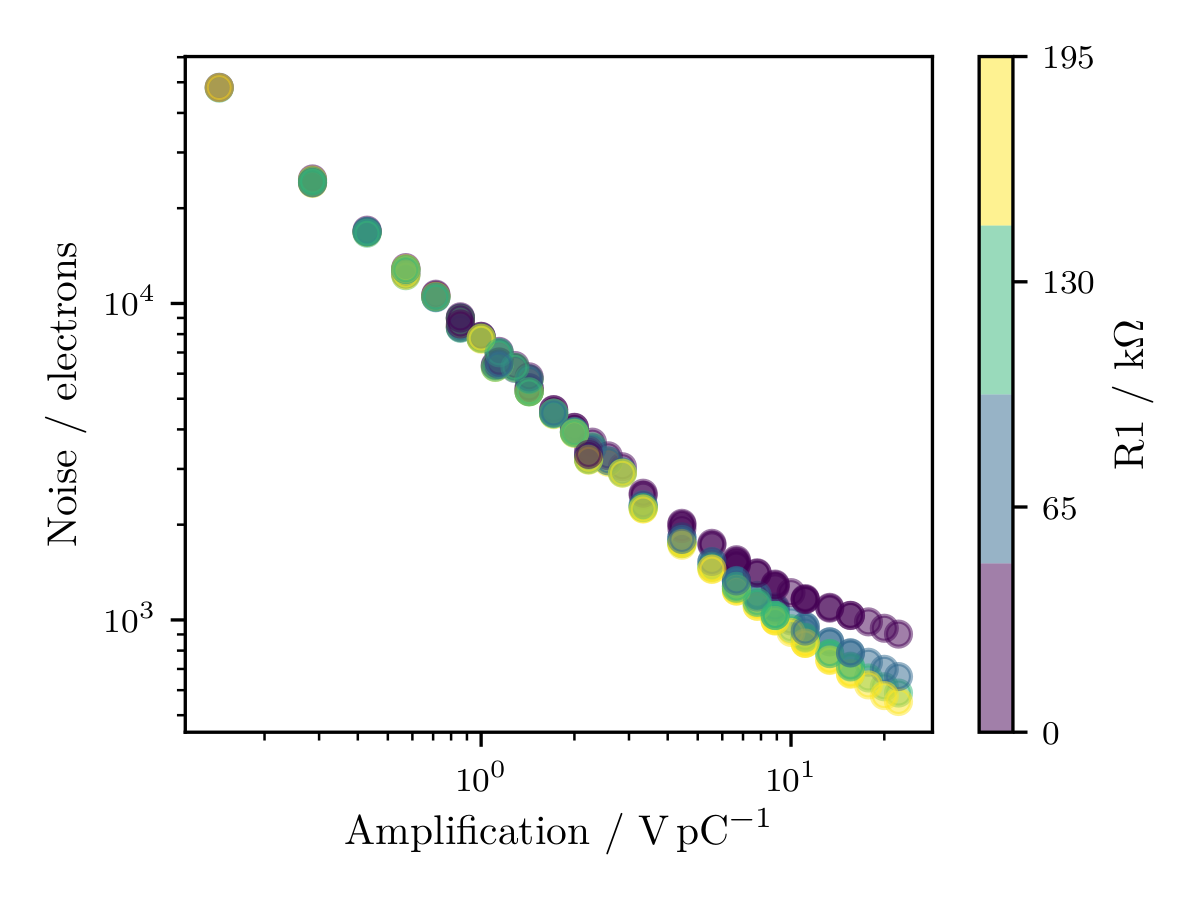}
    \caption{Measured noise of the AD8488 analog frontend as a function of gain and filter setting of the AD8488. At maximum amplification and lowest corner frequency of the built-in low-pass filter, the base noise of the AD8488 is less than \qty{600}{\elementarycharge}. }
    \label{fig:noise}
\end{figure}

After manufacturing of the prototype system shown in \cref{fig:foto}, first validation measurements were carried out. One crucial parameter is the noise of the analog front ends. This noise is highly dependent on the configuration of the AD8488 front-end. The AD8488 converts the input charge to an output voltage and enables selecting a wide range of amplification factors  \cite{analog-devicesDataSheetAD84882012}. Further, the AD8488 is equipped with a configurable low-pass filter to reduce the contribution of the high-frequency noise from the input amplifier on the output signal \cite{analog-devicesDataSheetAD84882012}. Enabling low-pass filtering comes at the cost of reduced measurement accuracy. Measurement results for the noise as a function of the charge to voltage conversion factor and of the low-pass filter are shown in \cref{fig:noise}. The corner frequency of the low-pass filter is determined by the selected filter resistor (R1) and the hold capacitor, which is \qty{5}{\pico\farad}  at highest amplification \cite{analog-devicesDataSheetAD84882012}, resulting in a minimum corner frequency of \qty{163}{\kilo\hertz}. When configuring the low-pass filter at his corner frequency, the base-noise of the front end was confirmed to be less than \qty{600}{\elementarycharge}. 

\begin{figure}
    \centering
    \includegraphics[width=1\linewidth]{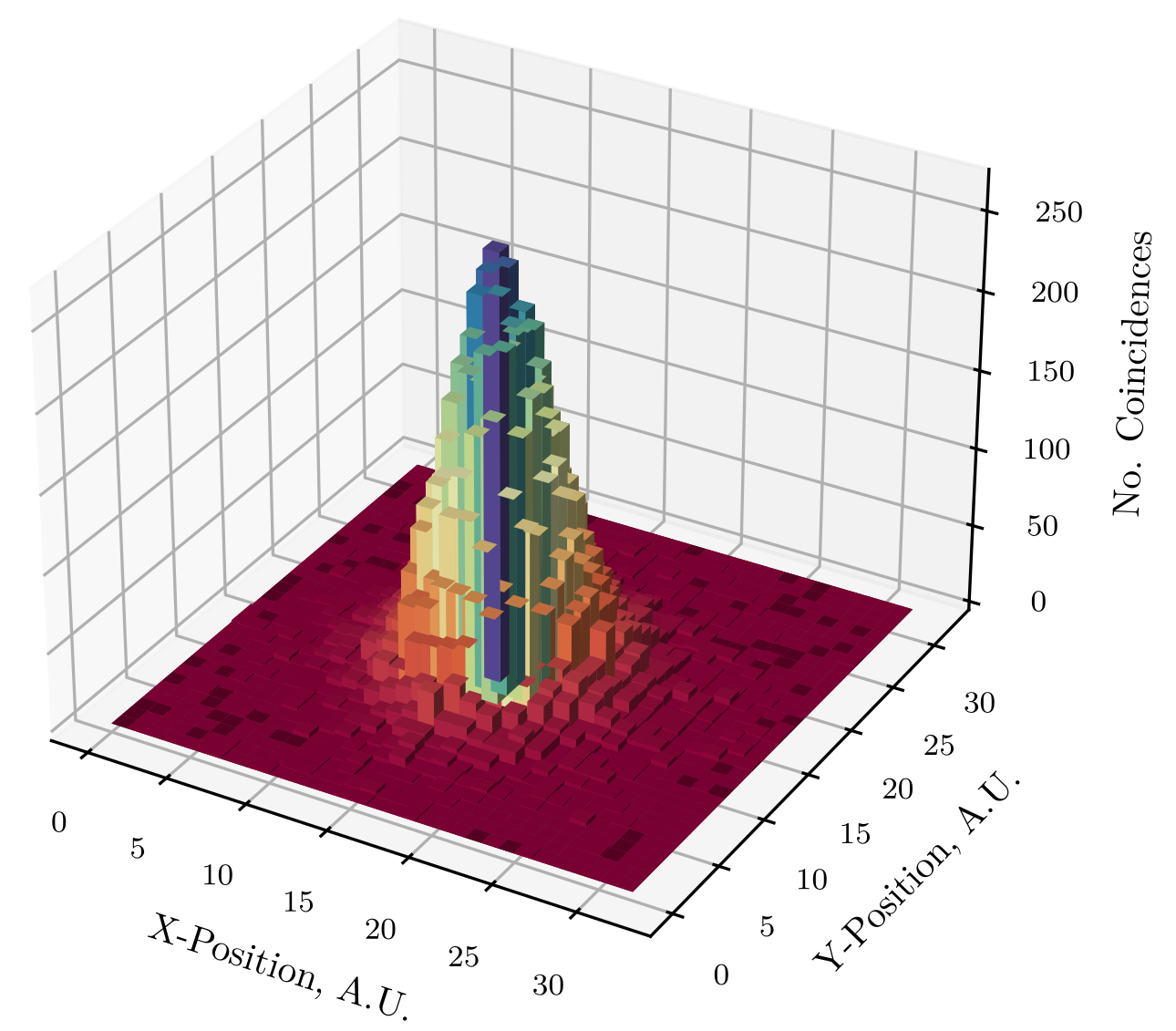}
    \caption{Beam profile of a 252.7 MeV proton beam reconstructed from coincidences of single particles. }
    \label{fig:beam_profile}
\end{figure}

As of writing, the designed SiC sensors are still in production. Therefore, Si strip sensors were employed to validate the performance of the beam monitor in beam tests. As a validation procedure, the monitor was configured in single-particle detection mode and exposed to a low flux beam. From the coincident detection of single particles on both detector planes, we reconstructed the beam profile. The result of such a reconstruction for a 252.7 MeV proton beam is shown in \cref{fig:beam_profile}.

\section{Conclusion and Outlook}

We developed a beam monitor capable of detecting single particles and clinical beams at the MedAustron cancer treatment facility in Wiener Neustadt. The monitor is capable of sampling at \qty{37}{\kilo\hertz} in synchronous mode and \qty{37}{\kilo\hertz} when interleaving measurements of odd and even strips. The design is based on our previous study \cite{waidSiCBasedBeam2024} and is built around the commercially available Analog Devices AD8488 charge integrator to read out a strip detector. The system is designed to use SiC sensors to enable DC-coupled charge readout. Given the early stage of SiC manufacturing technology and the expected low production yield, sensors were optimized for assembly from tiled elements. Correspondingly, guard rings were optimized for low area usage and high-voltage operation, using TCAD. The guard ring design enlarges the sensors by \qty{320}{\micro\meter}, keeping the dark area around the sensor low. A base noise of the readout system of \qty{600}{\elementarycharge} was measured. The beam monitor's functionality was validated by reconstruction of the beam profile of a 252.7 MeV proton beam.

The beam monitor was tested using Si sensors, and the production of SiC sensors is ongoing. However, the beam monitor is equipped with interfaces allowing it to work not only as a measurement instrument but also as an active control system. Once SiC sensors are available, we will use the newly developed system as a beam monitor and experiment with beam delivery algorithms. For this purpose, the monitor can communicate measurement data via SFP+ or send control signals via an I/O extender board.

\section{Acknowledgment}
This project has received funding from the Austrian Research Promotion Agency FFG, grant number 883652.

\bibliographystyle{elsarticle-harv} 
\bibliography{HEPHY-detector-dev}

\begin{thebibliography}{5}
\expandafter\ifx\csname natexlab\endcsname\relax\def\natexlab#1{#1}\fi
\providecommand{\url}[1]{\texttt{#1}}
\providecommand{\href}[2]{#2}
\providecommand{\path}[1]{#1}
\providecommand{\DOIprefix}{doi:}
\providecommand{\ArXivprefix}{arXiv:}
\providecommand{\URLprefix}{URL: }
\providecommand{\Pubmedprefix}{pmid:}
\providecommand{\doi}[1]{\href{http://dx.doi.org/#1}{\path{#1}}}
\providecommand{\Pubmed}[1]{\href{pmid:#1}{\path{#1}}}
\providecommand{\bibinfo}[2]{#2}
\ifx\xfnm\relax \def\xfnm[#1]{\unskip,\space#1}\fi
\bibitem[{{Analog-Devices}(2012)}]{analog-devicesDataSheetAD84882012}
\bibinfo{author}{{Analog-Devices}}, \bibinfo{year}{2012}.
\newblock \bibinfo{title}{Data {{Sheet AD8488}}}.
\bibitem[{Gaggl et~al.(2022)Gaggl, Bergauer, G{\"o}bel, Thalmeier, Villa and
  Waid}]{gagglChargeCollectionEfficiency2022}
\bibinfo{author}{Gaggl, P.}, \bibinfo{author}{Bergauer, T.},
  \bibinfo{author}{G{\"o}bel, M.}, \bibinfo{author}{Thalmeier, R.},
  \bibinfo{author}{Villa, M.}, \bibinfo{author}{Waid, S.},
  \bibinfo{year}{2022}.
\newblock \bibinfo{title}{Charge collection efficiency study on
  neutron-irradiated planar silicon carbide diodes via {{UV-TCT}}}.
\newblock \bibinfo{journal}{Nuclear Instruments and Methods in Physics Research
  Section A: Accelerators, Spectrometers, Detectors and Associated Equipment}
  \bibinfo{volume}{1040}, \bibinfo{pages}{167218}.
\newblock \DOIprefix\doi{10.1016/j.nima.2022.167218}.
\bibitem[{{Ulrich-Pur} et~al.(2021){Ulrich-Pur}, Adler, Bergauer, Burker,
  De~Franco, Guidoboni, Hirtl, Irmler, Kaser, Nowak, Pitters, Pivi,
  Prokopovich, Schmitzer and Wastl}]{ulrich-purCommissioningLowParticle2021}
\bibinfo{author}{{Ulrich-Pur}, F.}, \bibinfo{author}{Adler, L.},
  \bibinfo{author}{Bergauer, T.}, \bibinfo{author}{Burker, A.},
  \bibinfo{author}{De~Franco, A.}, \bibinfo{author}{Guidoboni, G.},
  \bibinfo{author}{Hirtl, A.}, \bibinfo{author}{Irmler, C.},
  \bibinfo{author}{Kaser, S.}, \bibinfo{author}{Nowak, S.},
  \bibinfo{author}{Pitters, F.}, \bibinfo{author}{Pivi, M.},
  \bibinfo{author}{Prokopovich, D.}, \bibinfo{author}{Schmitzer, C.},
  \bibinfo{author}{Wastl, A.}, \bibinfo{year}{2021}.
\newblock \bibinfo{title}{Commissioning of low particle flux for proton beams
  at {{MedAustron}}}.
\newblock \bibinfo{journal}{Nuclear Instruments and Methods in Physics Research
  Section A: Accelerators, Spectrometers, Detectors and Associated Equipment}
  \bibinfo{volume}{1010}, \bibinfo{pages}{165570}.
\newblock \DOIprefix\doi{10.1016/j.nima.2021.165570}.
\bibitem[{Waid et~al.(2024a)Waid, Gsponer, Burin, Gaggl, Thalmeier and
  Bergauer}]{waidSiCBasedBeam2024}
\bibinfo{author}{Waid, S.}, \bibinfo{author}{Gsponer, A.},
  \bibinfo{author}{Burin, J.}, \bibinfo{author}{Gaggl, P.},
  \bibinfo{author}{Thalmeier, R.}, \bibinfo{author}{Bergauer, T.},
  \bibinfo{year}{2024}a.
\newblock \bibinfo{title}{{{SiC}} based beam monitoring system for particle
  rates from {{kHz}} to {{GHz}}}.
\newblock \bibinfo{journal}{Journal of Instrumentation} \bibinfo{volume}{19},
  \bibinfo{pages}{C04055}.
\newblock \DOIprefix\doi{10.1088/1748-0221/19/04/C04055}.
\bibitem[{Waid et~al.(2024b)Waid, Maier, Gaggl, Gsponer, Sieberer, Babeluk and
  Bergauer}]{waidDetectorDevelopmentParticle2024}
\bibinfo{author}{Waid, S.}, \bibinfo{author}{Maier, J.},
  \bibinfo{author}{Gaggl, P.}, \bibinfo{author}{Gsponer, A.},
  \bibinfo{author}{Sieberer, P.}, \bibinfo{author}{Babeluk, M.},
  \bibinfo{author}{Bergauer, T.}, \bibinfo{year}{2024}b.
\newblock \bibinfo{title}{Detector development for particle physics}.
\newblock \bibinfo{journal}{e+i Elektrotechnik und Informationstechnik}
  \DOIprefix\doi{10.1007/s00502-023-01201-w}.

\end{thebibliography}

\end{document}